\documentclass[aps,prl,twocolumn,superscriptaddress,showpacs]{revtex4-2}

\usepackage{caption}
\usepackage{subcaption}
\usepackage{graphicx}
\usepackage[colorlinks=true, citecolor=blue, linkcolor=blue, urlcolor=blue]{hyperref}

\usepackage{bm}% bold math
\usepackage{color}
\DeclareTextSymbol{\degre}{OT1}{23}
\usepackage{ulem}
\usepackage{amsmath}
\usepackage{xcolor}
\usepackage{amssymb}

\begin{document}

\title{Anomalous pressure dependence of the bulk modulus and Yb valence in cubic YbPd}

\author{B. Tegomo Chiogo}
\email {bodry.tegomo\_chiogo@helmholtz-berlin.de}
\affiliation{Helmholtz-Zentrum Berlin für Materialien und Energie, Hahn-Meitner-Platz 1, Berlin D-14109, Germany}

\author{V. Balédent}
\affiliation{Université Paris-Saclay, CNRS, Laboratoire de Physique des Solides, 91405 Orsay, France}
\affiliation{Institut universitaire de France (IUF)}

\author{J.-P. Rueff}
\affiliation{Synchrotron SOLEIL, L'Orme des Merisiers, Départementale 128, 91190 Saint Aubin, France}
\affiliation{ Laboratoire de Chimie Physique-Matière et Rayonnement, CNRS, UMR 7614, Sorbonne Université, 4 Place Jussieu, 75252 Paris, France}

\author{E. Saïman }
\affiliation{Université Paris-Saclay, CNRS, Laboratoire de Physique des Solides, 91405 Orsay, France}

\author{V. Poree}
\affiliation{Synchrotron SOLEIL, L'Orme des Merisiers, Départementale 128, 91190 Saint Aubin, France}
\affiliation{Universit\'e de Rennes, CNRS, Institut des Sciences Chimiques de Rennes-UMR6226, 35042 Rennes, France}

\author{T. Schweitzer}
\affiliation{Universit\'e de Lorraine, CNRS, Institut Jean Lamour,
F-54000 Nancy,
France}

\author{D. Wong}
\affiliation{Helmholtz-Zentrum Berlin für Materialien und Energie, Hahn-Meitner-Platz 1, Berlin D-14109, Germany}

\author{C. Schulz}
\affiliation{Helmholtz-Zentrum Berlin für Materialien und Energie, Hahn-Meitner-Platz 1, Berlin D-14109, Germany}

\author{T. Mazet}
\affiliation{Universit\'e de Lorraine, CNRS, Institut Jean Lamour,
F-54000 Nancy,
France}

\author{A. Chainani}
\affiliation{National Synchrotron Radiation Research Center, Hsinchu Science Park, Hsinchu 30076, Taiwan}

\author{D. Malterre}
\affiliation{Universit\'e de Lorraine, CNRS, Institut Jean Lamour,
F-54000 Nancy,
France}

\author{K. Habicht}
\affiliation{Helmholtz-Zentrum Berlin für Materialien und Energie, Hahn-Meitner-Platz 1, Berlin D-14109, Germany}
\affiliation{Institut für Physik und Astronomie,
Universität Potsdam, D-14476 Potsdam, Germany}

\date{\today}

\begin{abstract}

We investigate the Yb valence instabilities in the strongly correlated YbPd compound using resonant X-ray emission spectroscopy as a function of pressure across the charge-order (CO) transition. At a low temperature (T = 30 K) in the CO phase, the Yb $4f$ valence remains nearly constant up to a pressure P$_L$ = 1.5 $\pm {0.2}$ GPa, and then increases gradually at higher pressures. In contrast, at room temperature in the normal phase, an anomalous decrease of the Yb $4f$ valence is observed, without any accompanying structural phase transition. This behavior is corroborated by a systematic pressure-dependent decrease of the unit-cell volume. Based on a Birch-Murnaghan analysis, the compressibility indicates hardening of the lattice with applied pressure up to a distinct kink seen at P$_K$ = 1.6 $\pm {0.2}$ GPa. In contrast, for P $>$ P$_K$, the Yb $4f$ valence saturates and the compressibility reveals a counterintuitive pressure-induced softening. The results show a minimum in the compressibility of YbPd (with $f^{0}$-$f^{1}$ hole-type mixed-valence), reminiscent of the maximum in compressibility seen in the $\gamma$-$\alpha$ first-order isostructural phase transition in cerium (with  $f^{0}$-$f^{1}$ electron-type mixed-valence).

\end{abstract}
%%%%%%%%%%%%%%%%%%%%%%%%%%%%%%%%%%%%%%%%%%%%%%%%%%%%%%%%%%%%%%%%%%%%%%%%%%%%%%

\maketitle
%%% Introduction %%%%%%%%%%%%%%%%%%%%%%%%%%%%%%%%%%%%%%%%%%%%%%%%%%%%%%%%%%%
%
%\section{INTRODUCTION}

Cerium and ytterbium-based Kondo-lattice systems have attracted considerable attention over the past two decades owing to their outstanding physical properties, including unconventional superconductivity often associated with quantum phase transitions\cite{Hilbert2007, Gegenwart2008}. These original properties are due to the competition between the Ruderman-Kittel-Kasuya-Yosida (RKKY) interaction and the Kondo effect. These properties may be tuned through control parameters such as chemical doping, pressure, or magnetic field, and the competition between the two interactions is well summarized in the Doniach phase diagram \cite{DONIACH1977}. When the hybridization between the 4f states and the conduction band is weak, the RKKY interaction dominates. This leads to a stable trivalent 4f state and to the formation of large local magnetic moments, which order magnetically at low temperatures. As the hybridization strength increases, the Kondo interaction becomes dominant: the 4f magnetic moments are screened by the conduction electrons, and the valence deviates from the trivalent state.

\begin{figure}[htb]
\vspace{-0.2cm}
\begin{center}
\scalebox{0.36}{\includegraphics{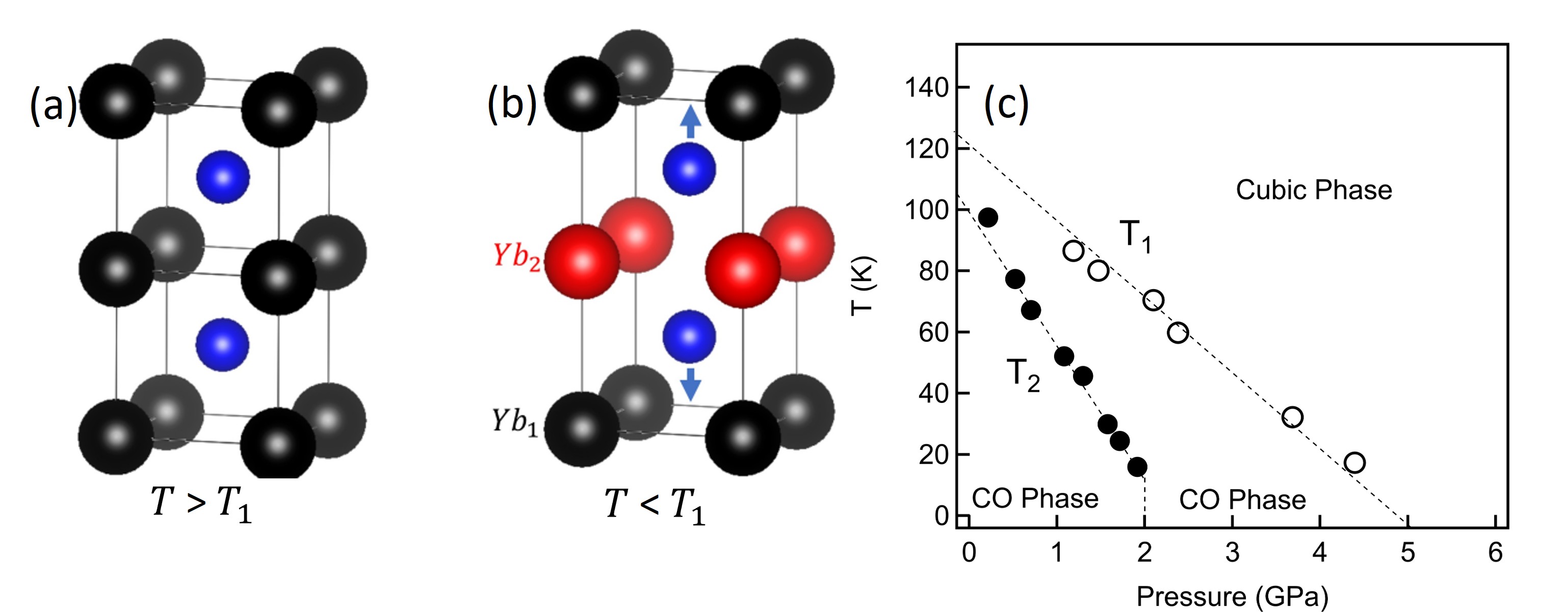}}
\end{center}\vspace{-0.5 cm}
\caption{\label{crystal_structure}Crystal structure of YbPd at ambient pressure (a ) above $T_1$ = 130 K (cubic phase) and (b) below $T_1$
(tetragonal phase). YbPd shows only one type of Yb-site in the normal cubic phase, but shows two types of Yb-sites in the tetragonal CO phase. (c) P-T phase diagram reported in Ref \cite{Miyake2012, Oyama2019}.} \vspace{-0.0 cm}
\end{figure}

 The strongly correlated YbPd material is a unique Yb compound that exhibits both, intermediate valence behavior and charge ordering (CO). It crystallizes in the cubic CsCl structure at room temperature and exhibits an antiferromagnetic phase transition at $T_N$ = 1.9 K due to the competition between the RKKY interaction and the Kondo effect \cite{Pott}.  Upon cooling from room temperature to $T_N$, YbPd undergoes two transitions. Below $T_1$ = 130 K, the cubic structure with a unique Yb site transforms into a tetragonal structure \cite{Bonville, Hasegawa2011} with two distinct Yb sites that form an incommensurate CO, which becomes a commensurate CO below $T_2$ = 105 K (Fig. \ref{crystal_structure}a, b). At low temperature, a trivalent magnetic Yb$^{3+}$ ion and strongly intermediate-valent nonmagnetic Yb$^{2.6+}$ ion are arranged alternatively along the $c$ axis, which has also been termed valence order \cite{Sugishima2010, Tokiwa2011,Takahashi,Shiga}. The CO/valence order in YbPd is very intriguing since YbPd shows metallic electrical resistivity from above $T_1$ down to low temperature. Accordingly, many studies have been devoted to understanding the magnetic ordering in the intermediate valence YbPd compound as well as the origin of the CO transition \cite{Pott,Takahashi,Mitsuda,Miyake2012,Miyake2013,Tsutsui2020}. Recently, the Kondo effect has been proposed as a mechanism linking the structural transition, the mixed valence, and the zero thermal expansion in the CO phase \cite{liao2022}.

 Pressure-dependent ac-calorimetry and electrical resistivity measurements revealed another peculiar behavior of the YbPd compound. Unlike typical ytterbium compounds, where pressure usually stabilizes magnetic ordering, in YbPd the magnetic transition temperature $T_N$ decreases with increasing pressure and is discontinuously suppressed at P$_C$ = 1.9 GPa \cite{Sugishima2010, Miyake2012, Miyake2013}. Both the CO transitions at $T_1$ and $T_2$ also decrease monotonically with increasing P (Fig. \ref{crystal_structure}c).
 A precise understanding of the exotic properties of YbPd requires detailed knowledge of the temperature and pressure dependence of the Yb valence. While temperature-induced valence changes have been investigated using high-energy spectroscopic techniques \cite{Pott,Domke,liao2022}, the pressure dependence of the valence evolution in YbPd has not yet been reported.

 In this study, we investigate the pressure dependence of the Yb valence in YbPd under high pressure at temperatures below and above the CO transition by means of High energy resolution Fluorescence detected X-ray absorption spectroscopy (HERFD-XAS) and Resonant X-ray Emission Spectroscopy (RXES) at the Yb L$_3$ edge (see supplementary materials (SM)\cite{SM} for experimental details), which are sensitive probes of the 4f valence state owing to their superior spectral resolution \cite{Dallera2002,Kummer2011,kummer2018}. To study the interplay between the Yb valence and the CO, we measured the pressure (P) dependence of the Yb valence below the CO transitions in the tetragonal phase (at T = 30 K) and in the cubic phase (at room temperature T = 300 K). At low temperature (T = 30 K), the valence is constant up to P$_L$ $\sim$ 1.5 $\pm {0.2}$ GPa but it increases at higher P. Remarkably, the measurements at room temperature revealed a significant decrease of the Yb valence at low pressures P$<$ P$_K$, followed by a nearly constant mixed valence at higher P. This is accompanied by a kink at P = P$_K$ in the P-dependent unit cell volume. Moreover, while the compressibility $\beta$ of most materials generally increases with P, YbPd exhibits a minimum in $\beta$ vs. P at P = P$_K$. This behavior is reminiscent of the changes in $\beta$, albeit with a maximum instead of minimum, at the $\gamma$-$\alpha$ first-order isostructural phase transition in cerium metal \cite{bridgman1927, Lipp2008,Lipp2017}.

\begin{figure}[htb]
\vspace{-0.2cm}
\begin{center}
\scalebox{0.4}{\includegraphics{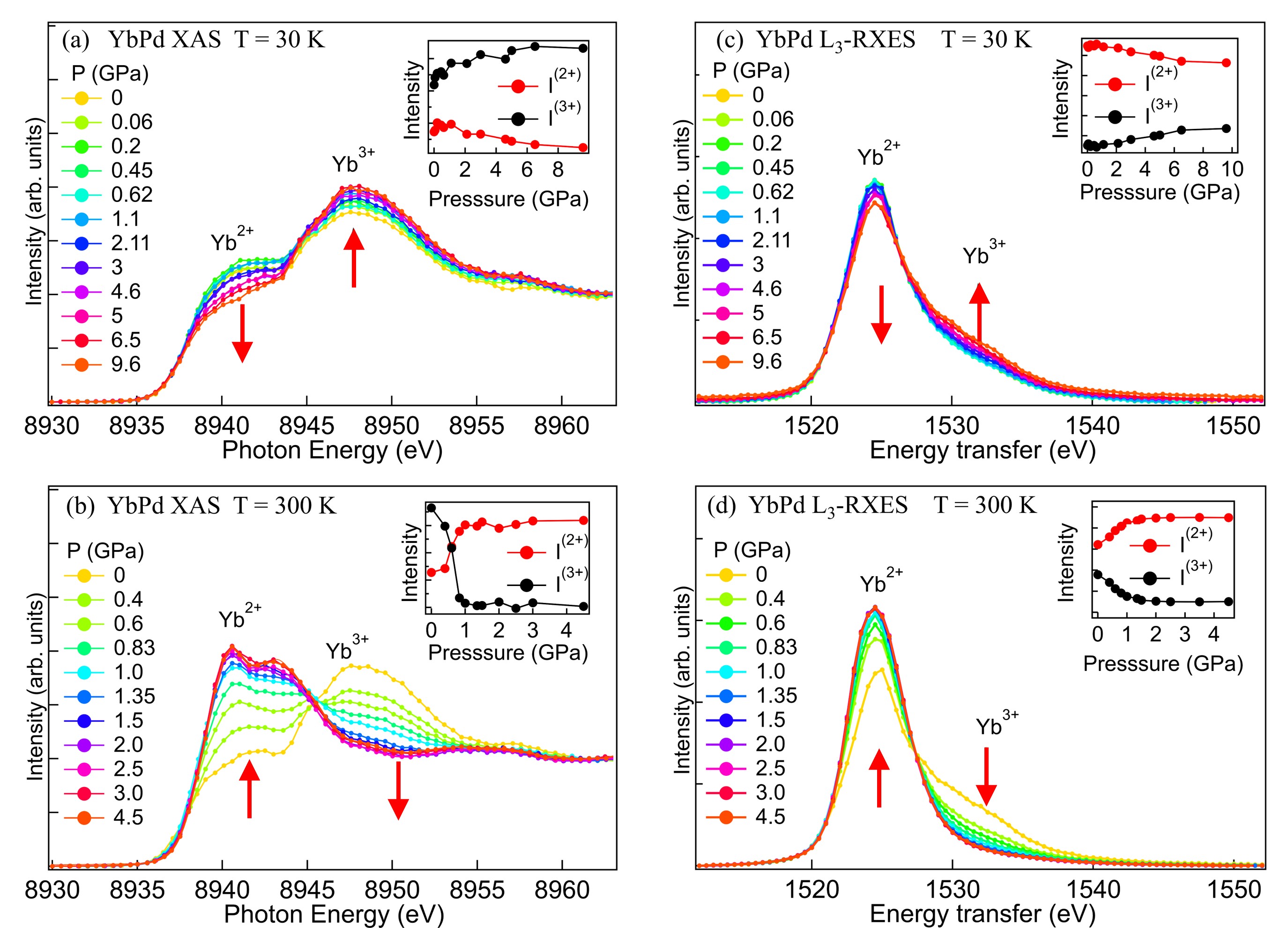}}
\end{center}\vspace{-0.5 cm}
\caption{\label{YbPd_XAS_XES_Pdep}Yb L$_3$-XAS and RXES spectra of YbPd as a function of pressure at  (a) $T$ = 30 K and (b) $T$ = 300 K. The incident photon energy of the RXES spectra was set to the maximum of the Yb$^{2+}$ resonance at the Yb L$_3$ absorption edge. The inset shows the evolution of the integrated intensities of the Yb$^{2+}$ and Yb$^{3+}$ final states with pressure.} \vspace{-0.0 cm}
\end{figure}

%\section{RESULTS}

Fig.~\ref{YbPd_XAS_XES_Pdep}a shows the pressure-dependent Yb-L$_3$ HERFD-XAS spectra of YbPd measured at $T = 30$ K. The spectra were recorded by scanning the incident photon energy across the Yb L$_3$ absorption edge while monitoring the scattered intensity of the Yb L$_{\alpha1}$ fluorescence line ($ 2p^5 3d^{10} \rightarrow 2p^6 3d^9, h\nu_{\text{out}} = 7417 \, \text{eV}$). The HERFD-XAS spectral shape is sharper than the standard XAS spectrum because the broadening is determined by the longer lifetime of the shallower 3$d$ hole, rather than by that of the deep 2$p$ hole \cite{Berman1991}. The spectra exhibit two peaks at around 8940 eV and 8947 eV corresponding to the transition from the initial intermediate valence ground state $\big|g\big\rangle = a\big|f^{13}\big\rangle + b\big|f^{14}\big\rangle$ to a final state with mainly Yb$^{2+}$ ($ 2p^6 4f^{14} \rightarrow 2p^5 4f^{14} 5d^1$) and Yb$^{3+}$ ($ 2p^6 4f^{13} \rightarrow 2p^5 4f^{13} 5d^1$) characters. This intermediate valence character at low temperatures is consistent with early XAS at Yb $L_3$ edge (average valence $v =$ 2.80) \cite{Pott} as well as soft x-ray photoemission   ($v =$ 2.75)\cite{Domke} and hard x-ray ($v =$ 2.79)\cite{liao2022} photoemission measurements. The Yb$^{2+}$ structure is split into two peaks separated in energy by $\sim$2.3 eV. This splitting is attributed to the 5d crystal field effect \cite{Yamaoka2008,kotani2012}. With increasing pressure, a clear spectral-weight transfer from Yb$^{2+}$ to Yb$^{3+}$ is observed. In contrast, at room temperature, the Yb$^{3+}$ decreases and is transferred to Yb$^{2+}$ with increasing pressure as shown in  Fig.~\ref{YbPd_XAS_XES_Pdep}b.

\begin{figure*}[htb]
\vspace{-0.2cm}
\begin{center}
\scalebox{0.42}{\includegraphics{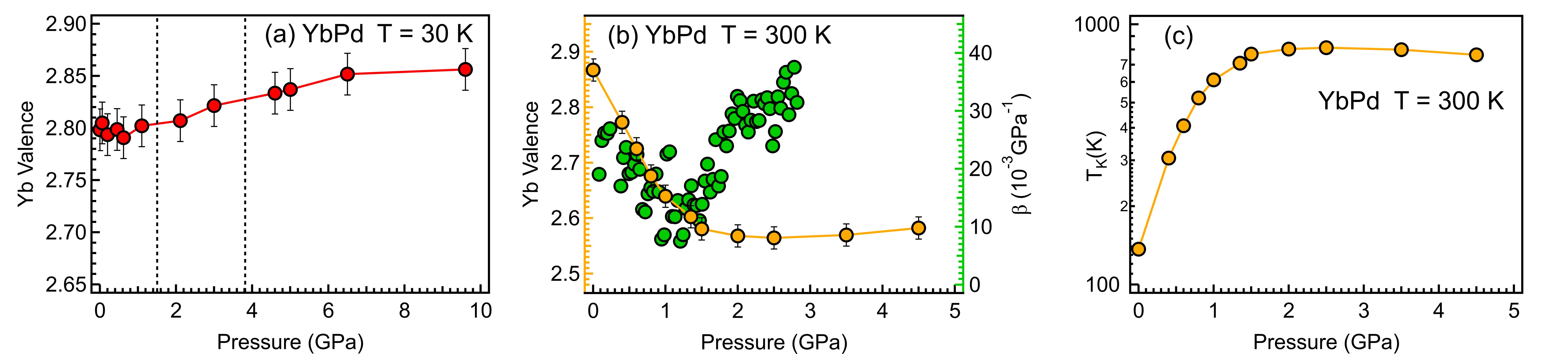}}
\end{center}\vspace{-0.5 cm}
\caption{\label{valence_YbPd_XES_Pdep_300K_30K}(a) P dependence of the Yb valence at T = 30 K. (b) P dependence of the Yb valence (left scale) and  the compressibility $\beta$ (right scale) at room temperature. (c) P dependence of the Kondo temperature at 300 K.} \vspace{-0.0 cm}
\end{figure*}

Extracting valence from XAS spectra is hampered by uncertainty in the edge-step position and background due to the edge-to-continuum jump. RXES is a more efficient way, thanks to its resonant character, which allows for the enhancement of either the Yb$^{2+}$ or Yb$^{3+}$ configuration, providing an accurate determination of the Yb valence \cite{Kummer2011, Mazet2013, kummer2018,Gannon2018}.
The pressure-dependent Yb L$\alpha_{1}$ RXES spectra measured with the incident photon energy tuned at the Yb$^{2+}$ emission line of the absorption edge for $T = $ 30 K and $T = $ 300 K are shown in Figure \ref{YbPd_XAS_XES_Pdep}c and \ref{YbPd_XAS_XES_Pdep}d. The spectra are plotted as a function of the transferred energy $E_t$ and normalized to the area under the curve to highlight the spectral changes. The RXES spectra exhibit two structures corresponding to Yb$^{2+}$ ($ 2p^6 4f^{14} \rightarrow 2p^5 4f^{14} 5d^1 \rightarrow 2p^6 3d^9 4f^{14}$) and Yb$^{3+}$ ($ 2p^6 4f^{13} \rightarrow 2p^5 4f^{13} 5d^1 \rightarrow 2p^6 3d^9 4f^{13}$) final states and similar spectral changes with pressure are observed, as in the XAS spectra. 

Let's now turn to the quantitative analysis of the Yb valence as a function of pressure. The spectra are analyzed using phenomenological fitting with a standard pseudo-Voigt function for each valence state (as detailed in the supplementary materials). The Yb valence $v$ is quantitatively determined from the RXES spectra using the expression \cite{Kummer2011,kummer2018}, 
\begin{equation}
\begin{aligned}
v = 3 - \frac{I^{(2+)}}{I^{(2+)} + \frac{I^{(3+)}}{\zeta}},
\end{aligned}
\label{eq:Birch-Murnaghan}
\end{equation}
where $I^{(2+)}$ and $I^{(3+)}$ are, respectively, the integrated intensities of the peaks Yb$^{2+}$ and Yb$^{3+}$. The parameter $\zeta$ accounts for the reduced Yb$^{3+}$ spectral weight observed in RXES spectra, which depends on the incident photon energy. A value of $\zeta = 8.5\%$ yields a valence at ambient pressure consistent with that determined by hard X-ray photoemission spectroscopy ($v =$ 2.79 at 20 K) \cite{liao2022}.

%%%%%% 

%%%%%%%%%%%%%%%%%%%%%%%%%%%%%%%%%%%%%%%%%%%%%%%%%%%%%%%%%%%%%%%%%%%%%%%%%%%%%%%%%%%%%%
%%%%%%%%%%%%%%%%%%%%%%%%%%%%%%%%%%%%%%%%%%%%%%%%%%%%%%%%%%%%%%%%%%%%%%%%%%%%%%%%%%%%

Figs.~\ref{valence_YbPd_XES_Pdep_300K_30K}a and~\ref{valence_YbPd_XES_Pdep_300K_30K}b show the pressure dependence of the Yb valence obtained from RXES at $T = $ 30 K and $T$ = 300 K, respectively. In Fig.~\ref{valence_YbPd_XES_Pdep_300K_30K}a, the dotted vertical lines mark the pressure induced transition as found from earlier work (as discussed in \cite{Miyake2012} and refs. therein), the commensurate CO state is stable up to P$_L$ $\sim$ 1.5 $\pm {0.2}$ GPa, where it transforms into the incommensurate CO and the cubic phase is subsequently restored for P higher than 3.7 GPa at T = 30 K. The mean Yb valence increases globally  upon increasing pressure, as expected. It is almost P independent in the commensurate CO state, increases throughout the incommensurate CO  range and gradually saturates in the normal cubic phase. In the low P quadratic commensurate CO phase, there are two Yb sites with valencies of $\sim$  3.0 and $\sim$ 2.6  for Yb$_1$ and Yb$_2$, respectively \cite{liao2022}, while in the cubic high P phase there is only a single Yb site with valence $\sim$2.85. Hence, while the pressure increase yields an unusual decrease of the Yb$_1$ valence, the mean valence does show a gradual increase from $v$ $\sim$2.80 to $\sim$2.85\cite{liao2022}, because a pressure increase leads to an increase of the Yb valence and the Yb$^{3+}$ ($f^{13}$) ion is smaller compared to that of the Yb$^{2+}$ ($f^{14}$) ion\cite{Dallera2003, Fernandez2012, Sato2014, Matsubayashi2015, Mazet2015, Eichenberger2020, Eichenberger2023, Tegomo2022, Frontini2022, Imura2023}.

At room temperature, we directly - and unexpectedly - observe an anomalous pressure-induced change in the Yb valence. The Yb valence first drops at low pressure (P $\leq$ 1.5 GPa), where it decreases from 2.87 to 2.58, before remaining almost constant upon further pressure increase. 
A similar pressure-induced anomalous valence reduction with pressure has recently been observed in YbCu$_{5}$ \cite{yamaoka2017}, YbCu$_{4.5}$\cite{Yamaoka2018}, and YbCu$_{6.5}$\cite{Yamaoka2024}. In YbCu$_{4.5}$\cite{Yamaoka2018} the Yb valence decrease is linked to a structural first-order transition, while in YbCu$_{5}$ \cite{yamaoka2017} and YbCu$_{6.5}$\cite{Yamaoka2024}, it has been suggested that this anomalous behaviour might be due to peculiar features of the density of states near the Fermi level.

%\section{DISCUSSION}

The pressure dependence of the Kondo temperature T$_K$ in Ce- and Yb-based heavy-fermion compounds has been studied theoretically within the framework of the single impurity Anderson model (SIAM) \cite{Goltsev2005}. It was shown that for Ce compounds, T$_K$ increases monotonically, whereas for Yb compounds it decreases at low pressure, shows a broad minimum, and then increases with further increases in pressure. At room temperature, YbPd might be located near the broad minimum. It is well known that, in Yb-based compounds, the Kondo temperature \(T_K\) scales with the occupation of the \(4f\) hole number $n_h$ (the Yb valence $v = 2 + n_h$). By combining the Kondo energy scale \(T_K\), extracted from thermodynamic measurements, with the \(4f\)-state occupation determined by RIXS experiments, Kummer \textit{et al.} \cite{kummer2018} recently showed that this scaling follows a universal power law,
$3 - v = a T_K^{n},$ with \(n = 2/3\) and \(a = 1/200\). We employed this power law to estimate the pressure dependence of the Kondo temperature for the single Yb site at room temperature, as shown in Fig.~\ref{valence_YbPd_XES_Pdep_300K_30K}c. The Kondo temperature increases markedly from 137~K to 768~K at  P$_K$ $\sim$ 1.6 $\pm {0.2}$ GPa and then remains nearly unchanged upon further increase in pressure.

\begin{figure}[htb]
\vspace{-0.2cm}
\begin{center}
\scalebox{0.28}{\includegraphics{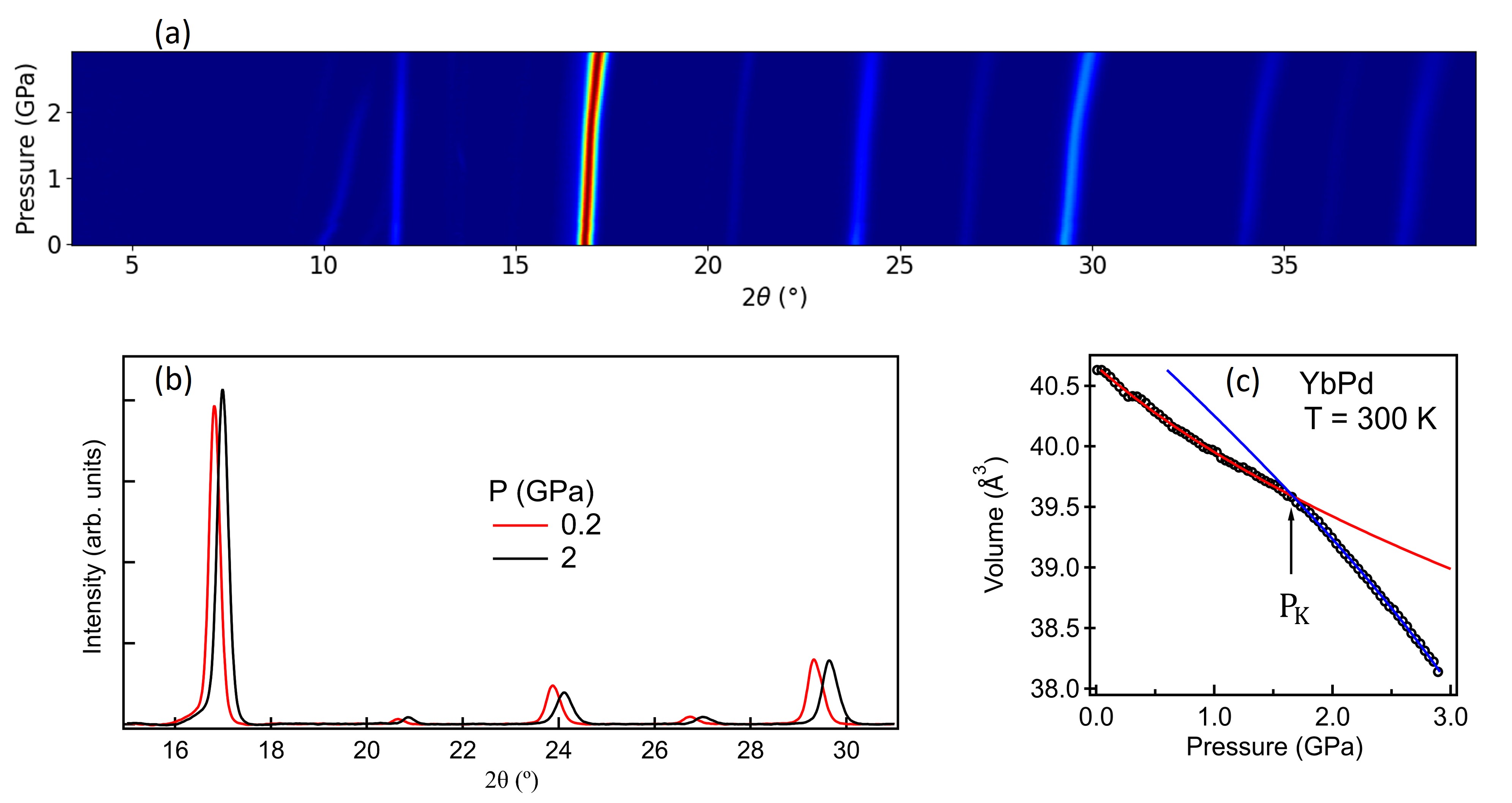}}
\end{center}\vspace{-0.5 cm}
\caption{\label{XRD_map} (a) Contour map of X-ray diffraction intensities collected in the pressure range 0–3 GPa. (b) XRD patterns at P = 0.2 GPa and 2 GPa. (c) Pressure dependence of the volume. The blue and red curves are the third-order Birch-Murnaghan’s equation of state for P $\leq$ P$_K$ and P $\geq$ P$_K$ respectively. The arrow indicates the pressure P$_K$ where the kink is observed.} \vspace{-0.0 cm}
\end{figure}

We performed room temperature P dependent X-ray diffraction (XRD) measurements (see SM\cite{SM} for experimental details) to investigate how the Yb valence variation couples to the crystal structure.
Fig.~\ref{XRD_map}a shows the contour map of X-ray diffraction intensities as a function of pressure; no structural phase transition is detected up to 3 GPa, as shown in Fig. \ref{XRD_map}b. Fig.~\ref{XRD_map}c shows the pressure dependence of the unit cell volume extracted from the XRD data. The unit cell volume decreases with pressure with a kink near P$_K$ $\sim$ 1.6 $\pm {0.2}$ GPa. 

At low P, the compressibility $\beta = -\frac{1}{V} \frac{\partial V}{\partial p}$ is found to decrease with increasing pressure as is typical for solids \cite{Ledbetter1973}. On increasing P further, it exhibits a minimum near P$_K$ $\sim$1.6 $\pm {0.2}$ GPa, the pressure at which the Yb valence saturates, and beyond P$_K$, the compressibility increases at higher pressures indicating an unusual behavior not reported in Yb-based mixed valent materials to date (Fig~\ref{valence_YbPd_XES_Pdep_300K_30K}b, right scale). This indicates a change of stiffness which we quantify using a Birch-Murnaghan analysis in the following.
We extracted the bulk modulus $K_0$ (the reciprocal of the compressibility), which characterizes the compression behavior under hydrostatic pressure, and its pressure derivative $K^{'}_0 =  \frac{\partial K_0}{\partial p}$ that defines the change in stiffness using the standard third-order Birch-Murnaghan equation of state \cite{Birch1947}:
\begin{equation}
\begin{aligned}
P (V) = \frac{3}{2} K_0 
&\left[ \left( \frac{V_0}{V} \right)^{\frac{7}{3}} 
- \left( \frac{V_0}{V} \right)^{\frac{5}{3}} \right] \\
&\times \left[ 1 + \frac{3}{4} (K_0' - 4) 
\left( \left( \frac{V_0}{V} \right)^{\frac{2}{3}} - 1 \right) \right]
\end{aligned}
\label{eq:Birch-Murnaghan}
\end{equation}
where P and V are the pressure and volume, respectively,
and V$_0$ and K$_0$ are the values of V and K at zero pressure. The data for the V(P) functions, illustrating the fitted equation of state, are shown in Figure \ref{XRD_map}c. We obtain the parameter of $K^{'}_0$ = 22.3, $K_0$ = 47.18 GPa, $V_0$ = 40.66 \AA$^3$ for P $<$ P$_K$ and $K^{'}_0$ = -6.1, $K_0$ = 38.32 GPa,  $V_0$ = 39.6 \AA$^3$ for P $>$ P$_K$. Remarkably, we observe a sign change in the derivative of the bulk modulus $K^{'}_0$.

Negative values of $K'_0$ are quite rare and correspond to an effect known as \textit{pressure-induced softening} \cite{Tsiok1998,Drymiotis2004,Pantea2006,Fang2013,Fang2013_2,Fang2014,Cairns2015,Wei2020}, a counter-intuitive phenomenon in which a material becomes increasingly compressible under pressure. 
Pressure-induced softening has been observed in a few but very different materials, including cerium metal\cite{bridgman1927}, the amorphous silica \cite{Tsiok1998}, ZrW$_2$O$_8$\cite{Pantea2006}, Zn(CN)$_2$\cite{Fang2013} and ScF$_3$\cite{Wei2020}. Among these materials, cerium exhibits strong similarities with YbPd. Indeed, cerium undergoes an isostructural face-centered cubic (fcc) $\gamma$-$\alpha$ phase transition, accompanied by a volume collapse of approximately 15\% at room temperature under pressure. The bulk modulus of cerium exhibits a minimum at the 
$\gamma$-$\alpha$ volume-collapse transition that persists even above the critical endpoint in the supercritical state \cite{bridgman1927,Jeong2004, Lipp2008, Decremps2011, Lipp2017}. Moreover, the phase transition in Ce is marked by a pronounced increase in the Kondo temperature, occurring concomitantly with the volume collapse under pressure. The $f$ electrons play a central role in this transition. As shown in Fig.~\ref{valence_YbPd_XES_Pdep_300K_30K}c, YbPd likewise exhibits a rapid increase in the Kondo temperature at low pressures up to P$_K$ - at which the compressibility reaches a minimum - and then becomes nearly constant above P$_K$. The sign change of  $K^{'}_0 $ is reminiscent to that observed in Ce. 
Both the pressure-temperature phase diagram and the temperature dependence of the bulk modulus of cerium have been successfully reproduced within the Kondo volume collapse model \cite{Allen1982,LAVAGNA1982,Allen1992} in which a rapid pressure-induced increase of the Kondo temperature drives the transition. Our experimental findings identify strongly correlated YbPd as an interesting system to test the applicability of the Kondo volume collapse model in describing pressure-induced changes resulting in a pronounced hardening-to softening transition.

In conclusion, we have investigated the pressure dependence of the Yb valence in YbPd using Yb-L$_3$ resonant X-ray emission spectroscopy. Our RXES data in the low-temperature charge ordering phase show a gradual increase in the mean valence with increasing pressure. In contrast, the RXES data at room temperature in the normal phase showed that the Yb valence decreases from 2.87 to 2.58 at 1.6 GPa and subsequently saturates with further increase of the pressure. This unexpected and unusual Yb valence evolution is coupled with an uncommon hardening-to-softening phenomenon. This behavior is somewhat surprising and has not been reported in previous studies, likely because this region of the pressure-temperature  phase diagram of YbPd (as well as other Yb-based mixed valent systems) had remained largely unexplored. Most of the earlier P-dependent studies focussed on investigating  the low temperature behavior. By analogy with Ce, this pronounced minimum in the pressure dependence of the compressibility suggests proximity to a first-order transition in the ($P$,$T$) phase diagram of YbPd.

%%%%%%%%%%%%%%%%%%%%%%%%%%%%%%%%%%%%%%%%%%%%%%%%%%%%%%%%%%%%%%%%%%%%%%%%%%%%%%%%%%%%%%%%%%
%%%%%%%%%%%%%%%%%%%%%%%%%%%%%%%%%%%%%%%%%%%%%%%%%%%%%%%%%%%%%%%%%%%%%%%%%%%%%%%%%%%%%%%%%%%%%%

\section{ACKNOWLEDGMENTS}
We acknowledge the French Synchrotron facility SOLEIL (Saint-Aubin, France) for the allocated beam time (exp. n° 20241507). This work was supported by the Paris Ile-de-France Region in the framework of DIM MaTerRE (project DAC-VX).

\bibliography{Ref.bib}

\end{document}